\documentclass[pdflatex,sn-mathphys-num]{sn-jnl}
\usepackage{textcomp}

\usepackage{tabularx}
\usepackage{multirow}
\usepackage{amsmath}
 \usepackage{booktabs}
\usepackage{longtable}
\usepackage{pdflscape}
\usepackage{array}
\usepackage{placeins}
\usepackage{eurosym}

\usepackage{listings}
\usepackage{xcolor}

\definecolor{codegreen}{rgb}{0,0.6,0}
\definecolor{codegray}{rgb}{0.5,0.5,0.5}
\definecolor{codepurple}{rgb}{0.58,0,0.82}
\definecolor{backcolour}{rgb}{0.95,0.95,0.92}

\lstdefinestyle{mystyle}{
    backgroundcolor=\color{backcolour},   
    commentstyle=\color{codegreen},
    keywordstyle=\color{magenta},
    numberstyle=\tiny\color{codegray},
    stringstyle=\color{codepurple},
    basicstyle=\ttfamily\footnotesize,
    breakatwhitespace=false,         
    breaklines=true,                 
    captionpos=b,                    
    keepspaces=true,                 
    numbers=left,                    
    numbersep=5pt,                  
    showspaces=false,                
    showstringspaces=false,
    showtabs=false,                  
    tabsize=2
}

\newcommand{\safeincludegraphics}[2][]{%
  \IfFileExists{#2}{\includegraphics[#1]{#2}}{%
    \fbox{\parbox[c][0.3\textheight][c]{0.95\linewidth}{\centering Figure placeholder: \texttt{#2}}}%
  }%
}

\begin{document}

\title[pySpainMobility]{pySpainMobility: Unlocking Spanish Open Mobility Data for Spatial Inequality Research}

\author*[1,2]{\fnm{Ciro} \sur{Beneduce}}\email{cbeneduce@fbk.eu}
\author[3]{\fnm{Tania} \sur{Gull\'on Mu\~noz-Repiso}}\email{tgullon@transportes.gob.es}
\author[1]{\fnm{Bruno} \sur{Lepri}}\email{lepri@fbk.eu}
\author[1]{\fnm{Massimiliano} \sur{Luca}}\email{mluca@fbk.eu}

\affil*[1]{\orgdiv{MobS Lab, Bruno Kessler Foundation}, \orgaddress{\city{Trento}, \country{Italy}}}
\affil[2]{\orgdiv{Department of Computer Science, University of Trento}, \orgaddress{\city{Trento}, \country{Italy}}}
    \affil[3]{\orgdiv{Division of Transportation Studies and Technologies, Ministry of Transport and Sustainable Mobility of Spain}, \orgaddress{\city{Madrid}, \country{Spain}}}

\abstract{
Human mobility shapes access to resources, opportunities, and services, making movement data a powerful lens for studying spatial and social inequality. Yet despite the growing availability of official open mobility datasets, their research potential is rarely realized because the technical overhead of retrieving, harmonizing, and processing them often crowds out substantive analysis. To address this, we introduce \texttt{pySpainMobility}, a Python package that automates the retrieval and harmonization of Spain's open mobility data across spatial resolutions and demographic strata, streamlining national-scale, reproducible analysis.
Using the package, we study income-stratified mobility inequality across Spain's inter-province network, drawing on district-level origin-destination flows for four representative weeks spanning the seasons of 2023. We construct income-specific mobility layers and show that socioeconomic stratification is deeply embedded in the structure of the national mobility system: low-income mobility is disproportionately concentrated in a narrow set of destinations and shorter in spatial reach, while high-income groups access a broader and more distant hierarchy of destinations. Low- and high-income layers consistently follow weakly aligned destination hierarchies across seasons, indicating that income groups navigate distinct mobility geographies rather than a shared one at different volumes. We further show that destination provinces themselves differ systematically in the income composition of the travelers they receive, with several provinces attracting arrivals disproportionately skewed toward one income group relative to the national seasonal baseline. These results demonstrate how official open mobility data, combined with accessible tooling, can be operationalized to reveal spatial inequality as a structural property of national mobility networks.
}

\keywords{ Human mobility, Open mobility data, Multilayer networks, Spatial inequality, Income segregation, Python package}

\maketitle

\section{Introduction}

Understanding how people move within and between urban and rural areas is crucial for addressing a wide range of societal challenges. Human mobility data inform research and policy decisions across multiple domains, including transportation planning and city optimization~\cite{batty2013, mazzoli2019, beneduce2025urban}, epidemic modeling~\cite{eubank2004modelling, colizza2007modeling, colizza2008epidemic, perkins2014theory, luca2023crime, moreno2025critical}, inequalities and segregation~\cite{iyer2024mobility, toth2021inequality, iyer2025understanding, moro2021mobility, park2024post}, and disaster response~\cite{bohorquez2009, bagrow2011, yabe2020quantifying, yabe2022mobile}. In each of these settings, mobility is not only a measure of movement volume, but also a relational structure linking places, populations, and opportunities.

Despite the ubiquity of digital traces generated by mobile phones, GPS devices, and connected applications, access to reliable and representative mobility data remains limited~\cite{luca2021survey, luca2023crime, yabe2024enhancing}. High-resolution datasets are often proprietary, expensive, or subject to restrictive licensing agreements that limit their use in scientific research and public sector applications~\cite{yabe2024enhancing}. When data are publicly available, they are often fragmented, poorly documented, or difficult to integrate into scalable and reproducible workflows~\cite{yabe2024enhancing}. These barriers are especially important for nationwide mobility datasets, where repeated downloading, spatial harmonization, and processing of large daily files can become a substantial obstacle even when the data themselves are open.

To address part of this gap, the Spanish Ministry of Transport and Sustainable Mobility began releasing detailed daily mobility data derived from mobile phone signals~\cite{MITMA2023, gullon2022hermes, poncedeleon2022covidflowmaps, gullon2023movilidad, conesa2024mixture}. This initiative began in response to the COVID-19 pandemic, with an initial collection phase spanning from February 2020 to May 2021. A second release phase, starting in January 2022 and continuing to the present, reflects an ongoing effort to improve transparency and enable data-driven decision-making. The released datasets include aggregated origin-destination matrices, trip-count indicators, and overnight stay locations, available at three spatial aggregation levels: districts, municipalities, and greater urban areas. The data are updated daily and published with a short delay~\cite{MITMA2023, gullon2022hermes, poncedeleon2022covidflowmaps, gullon2023movilidad, conesa2024mixture}.

However, open publication alone does not guarantee scientific usability. Researchers and practitioners still need to identify the correct files, retrieve them repeatedly across dates and spatial scales, harmonize spatial identifiers, handle geometries, convert raw files into analysis-ready formats, and document the workflow in a reproducible way. These tasks are technical but consequential: without standardized access and preprocessing, analyses become harder to replicate, compare, and extend. This is particularly limiting for network-based mobility research, where origin-destination matrices must often be transformed into weighted graphs, stratified by population attributes, and analysed across multiple time periods.

The paper makes two contributions. First, it describes the design, data model, and functionality of \texttt{pySpainMobility}, showing how the package organizes the main spatial units and mobility data types released by the Ministry. Second, it demonstrates the analytical value of the package through a use case on income-stratified mobility inequality in Spain's inter-province mobility network. Using district-level origin-destination data for four representative weeks in 2023, we construct income-specific mobility layers and analyse how income groups differ in destination concentration, destination hierarchy, spatial reach, and destination-level income selectivity. The use case shows that the package enables more than data retrieval: it supports reproducible, national-scale, network-based analysis of social and spatial inequality using official open mobility data.

The rest of the paper is structured as follows. Section~\ref{sec:sources} reviews the broader landscape of mobility data sources. Section~\ref{sec:tessellations} introduces the study areas used in the Spanish datasets and explains how they are managed in the library. Section~\ref{sec:mobilitydata} details the structure and technical usage of the main data types accessible through the package. Section~\ref{sec:use_case} presents the income-stratified mobility use case, including network construction, measures, results, and discussion. Section~\ref{subsec:usecase_discussion} concludes with a summary of contributions, limitations, and directions for future work. Code, documentation, and notebooks to reproduce all analyses are available at \href{https://github.com/pySpainMobility}{github.com/pySpainMobility} and \href{https://pyspainmobility.github.io/pySpainMobility/}{pySpainMobility.github.io/pySpainMobility/}.

\begin{figure}[t]
    \centering
    \safeincludegraphics[width=\linewidth]{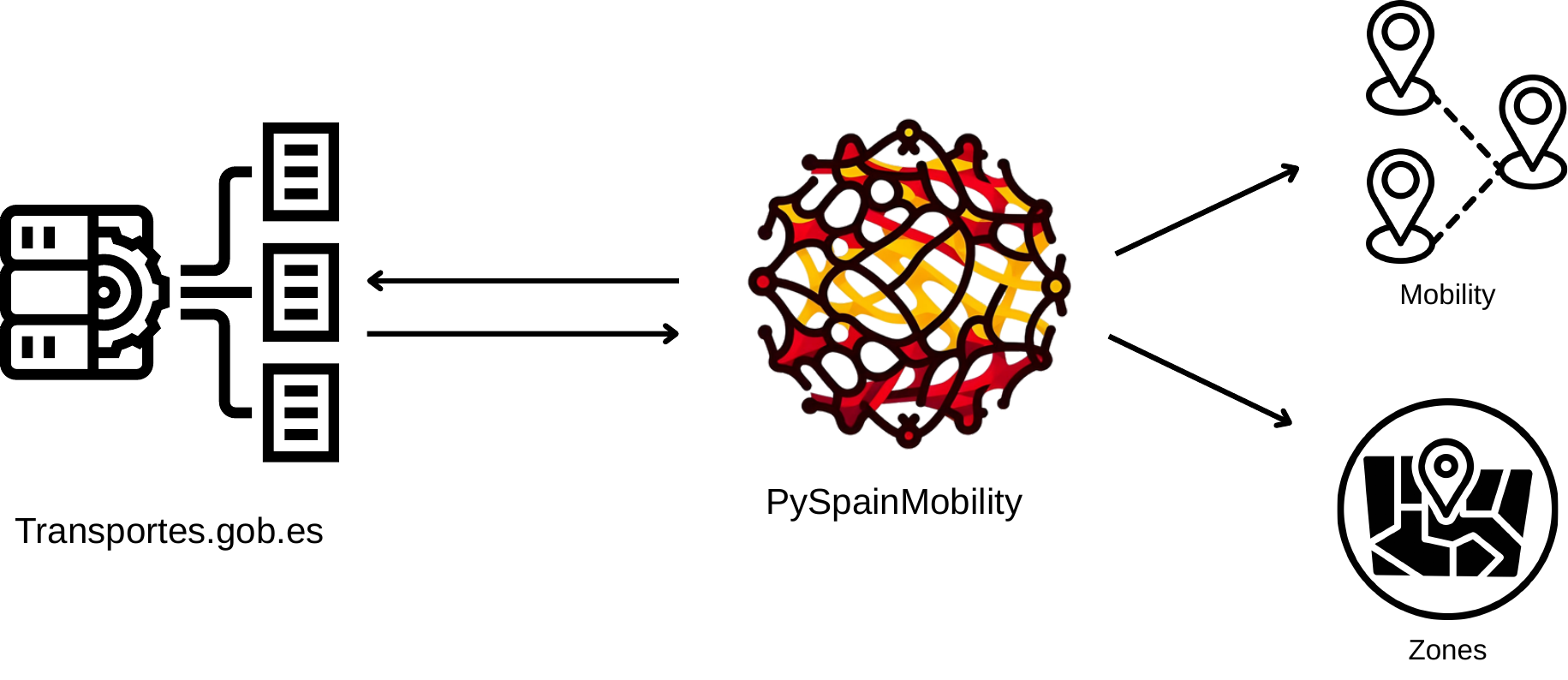}
    \caption{A simplified representation of how the library works. Data are 
    retrieved and processed directly from the website of the Ministry of 
    Transportation and Sustainable Mobility. Such data can be converted into 
    two different classes: \texttt{Mobility} and \texttt{Zones}.}
    \label{fig:library-schema}
\end{figure}

\section{Mobility Data Sources}
\label{sec:sources}

Before presenting the design of \texttt{pySpainMobility}, it is useful to situate
the Spanish datasets within the broader landscape of mobility data sources. Human
movement can be observed through several technologies, each with distinct
trade-offs in spatial accuracy, temporal resolution, population coverage, and
accessibility, and these trade-offs shape both the analyses that are feasible and
the reliability of their results.

The most widely used source is mobile network data, and in particular Call Detail
Records (CDRs), which log the time and serving cell tower of calls and text
messages. Because location is inferred from tower positions and recorded only when
an event occurs, CDRs are spatially coarse and temporally sparse, which limits
detailed trajectory reconstruction~\cite{gonzalez2008understanding, blondel2015survey, pappalardo2015returners}.
eXtended Detail Records (XDRs), generated by mobile-internet usage, mitigate this
sparsity by sampling locations more frequently and more
continuously~\cite{chen2019complete, luca2022modeling}.

GPS data collected through application SDKs or onboard navigation systems are far
more accurate in space and time, and several providers released such data during
the COVID-19 pandemic~\cite{chang2021mobility, aleta2022quantifying, moreira2013predicting}.
Their main drawbacks are limited transparency about how and from whom data are
collected, and known representativeness issues, since the social profiles of users
(e.g., age or income) are rarely disclosed. Aggregated indicators published by
technology firms such as Google and Facebook are broadly accessible and were widely
used to track behavioral change during lockdowns, but they offer lower spatial and
temporal granularity than raw traces~\cite{luca2023crime}.

Other sources are complementary. Geotagged social-media posts and check-ins are
sporadic and user-dependent but useful for tourism analysis and event
detection~\cite{llorente2015social}; credit-card transactions capture spatially
located spending and movement between residence, work, and
consumption~\cite{singh2015money}; satellite imagery supports urban-expansion
analysis when combined with mobility indicators~\cite{tatem2017worldpop}; and
wearables add physiological and activity data for travel-behavior and
health-related studies~\cite{wiedermann2022evidence}. Traditional census data
remain an essential baseline for population and commuting statistics but, given
their low update frequency, are poorly suited to real-time or event-driven
analysis~\cite{luca2023crime}.

Despite this proliferation of sources, a central limitation persists: the scarcity
of \emph{open and standardized} datasets that are comprehensive in scope and
reproducible in methodology~\cite{yabe2024enhancing}. Many datasets are restricted
to particular platforms, modes, or regions and rely on opaque preprocessing
pipelines whose undocumented choices (stop-detection thresholds, clustering
parameters, sampling criteria) materially affect the derived indicators and
undermine comparability across studies. The relative absence of shared benchmarks
further slows methodological progress compared with fields such as computer vision
or natural language processing~\cite{yabe2024enhancing}.

The datasets released by the Spanish Ministry of Transport and Sustainable Mobility
address part of this gap. They provide daily, nationwide indicators derived from
mobile-network data for more than 13 million users, at three levels of spatial
aggregation~\cite{MITMA2023, gullon2022hermes, gullon2023movilidad, conesa2024mixture}.
Data for 2020--2021 (version~1) are based on CDRs, whereas from 2022 onward
(version~2) a passive-measurement approach records devices attached to the network
even when inactive, not only during calls or data use, which substantially
improves data quality~\cite{pappalardo2021evaluation}. The provider's market share
exceeds 30\%, and correction procedures are applied to under-represented groups such
as children and the elderly, so that only limited segments of the population are
subject to self-selection bias~\cite{MITMA2023}. Open publication, however, does not by itself guarantee scientific usability: users must still locate and repeatedly download the correct files across dates and spatial scales, harmonize spatial identifiers, handle geometries, and convert raw files into analysis-ready formats. \texttt{pySpainMobility} responds to these challenges, a mission shared by parallel community efforts in R \cite{kotov2026spanishoddata}, through a Python-first workflow designed for scalable, reproducible, and analysis-ready access to Spanish mobility data by:

\begin{itemize}
\item generating \emph{analysis-ready mobility layers} by transforming raw demographic records into standardized, stratified origin–destination matrices, including dimensions such as income, age, and gender;
\item supporting \emph{national-scale longitudinal workflows} through an Apache Arrow backend and compressed Parquet caching, making large daily files and multi-season analyses tractable within standard Python environments;
\item unifying \emph{spatial and network pipelines} by linking mobility flows with official geometries across multiple territorial scales, from districts and municipalities to greater urban areas.
\end{itemize}

By integrating data retrieval, demographic stratification, efficient storage, and spatial interoperability, \texttt{pySpainMobility} provides a reproducible bridge between official mobility records and network-based urban research.

\section{Study Areas}
\label{sec:tessellations}
The Zones module in\texttt{ pySpainMobility }enables programmatic access to the spatial aggregations used in the official Spanish mobility datasets. These study areas divide the national territory into discrete, non-overlapping zones that serve as the fundamental spatial units for aggregating mobility indicators. The three available levels of granularity are districts (distritos), municipalities (municipios), and greater urban areas (grandes áreas urbanas).

These spatial definitions are essential for interpreting the data, and their configuration affects the analysis outcomes. For example, districts are the finest spatial resolution and tend to be larger in rural areas due to lower population density, while municipalities and greater urban areas are more heterogeneous and can better represent urban-rural transitions.

The Zones object supports flexible instantiation. Users must specify the desired spatial granularity using the zones parameter, which supports various aliases (e.g., `municipalities', `muni', `gau'). Additionally, the version parameter allows alignment with the temporal availability of the datasets: Version 1 covers data between February 2020 and May 2021 (excluding greater urban areas), while Version 2 includes data from January 2022 to June 2025. The study areas are retrieved from official sources and can be stored locally in GeoJSON format for further use.

To retrieve and load a study area as a GeoDataFrame, the user can run the code shown in Listing 1:

\begin{lstlisting}[language=Python, caption=Example of code to download and plot study areas]
from pySpainMobility import Zones

zones = Zones(zones='municipalities', version=2, output_directory='data')

gdf = zones.get_zone_geodataframe()

gdp.plot()
\end{lstlisting}

\begin{figure*}
    \centering
    \safeincludegraphics[width=1\linewidth]{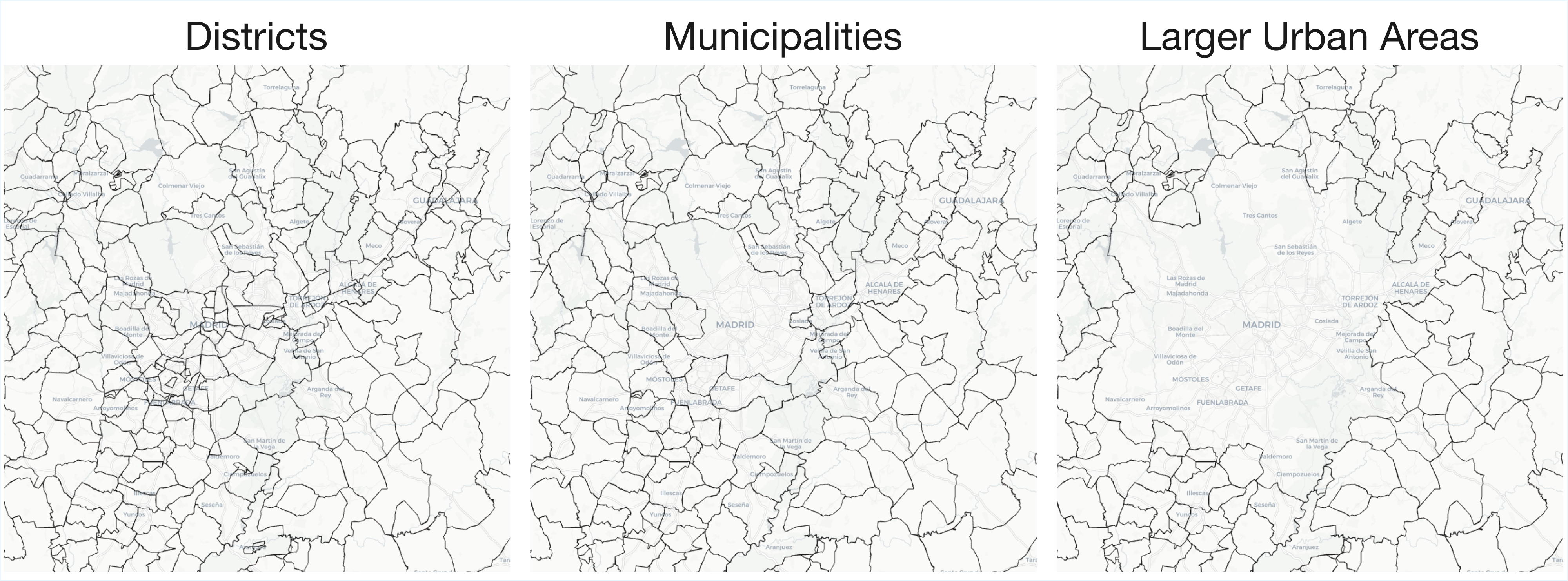}
    \caption{Example of different study areas over Madrid. On the left, the district aggregation, in the middle the aggregation of the municipalities, and on the right, the one of the larger urban areas.}
    \label{fig:tessellations}
\end{figure*}
This command downloads the relevant geometries and returns them as a GeoPandas object, allowing immediate spatial manipulation and visualization. Figure \ref{fig:tessellations} illustrates the differences between the three levels of aggregation and provides a zoomed-in view of the Madrid area, where spatial contrasts are most evident.

The average surface area of the zones varies across study area types. The average area size is 133.56 square kilometers for districts, 193.45 square kilometers for municipalities, and 242.79 square kilometers for Greater Urban Areas 

These values underline the substantial variation in spatial granularity across the country. In sparsely populated rural zones, even districts may span large areas, while in cities, the differences in resolution become more pronounced.

In addition to the geometries, the module provides access to a mapping table that allows interoperability between spatial aggregations and alignment with official statistics. This is particularly important when merging external data sources or when transitioning between datasets at different levels of aggregation.

To access the mapping table, the user can call the get\_zone\_relations() method, as shown in Listing 2:

\begin{lstlisting}[language=Python, caption=Example of code to obtain the dataframe with the zone identifiers to map different spatial aggregations.]
from pySpainMobility import Zones

zones = Zones(zones='municipalities', version=2, output_directory='data')

df = zones.get_zone_relations()
\end{lstlisting}

The returned DataFrame links identifiers across spatial schemes, allowing users to convert district codes to municipality or urban area codes, align mobility zones with official census references, and aggregate or disaggregate data across spatial hierarchies. This functionality ensures compatibility with external datasets and supports complex geographic analyses requiring hierarchical spatial logic.

\section{Mobility Data}
\label{sec:mobilitydata}
The Mobility module in\texttt{ pySpainMobility }provides a standardized interface for downloading and preprocessing the main datasets made available by the Spanish Ministry of Transport and Sustainable Mobility. These datasets are published under the national open data strategy and contain anonymized information derived from more than 13 million mobile phone users. The module allows access to daily mobility indicators in a programmatic and reproducible way, offering support for both historical (Version 1: February 2020 – May 2021 based on CDRs) and recent (Version 2: from January 2022 onward based on passive measurements) data collections.

Upon instantiation, the user specifies the version of the dataset, the spatial aggregation level (districts, municipalities, or greater urban areas), and the temporal range of interest. The zones parameter accepts various aliases (e.g., `distritos', `municipios', `gau') for user flexibility. The start date is mandatory, while the end date, if not specified, will be equal to the start date. All data are downloaded directly from the Ministry's open data portal and stored in Parquet format within the designated output directory. To keep national-scale processing tractable, the module parses the large daily files through an Apache Arrow backend by default, with an automatic fallback to pandas and optional Dask support for out-of-core workloads, and caches downloaded data as compressed Parquet to avoid repeated transfers.

The main functionalities of the module are described below.

\subsection{Origin–Destination Matrices}
Origin–destination (OD) matrices represent the volume and characteristics of daily flows between pairs of geographic zones. Each record captures aggregated movements from an origin to a destination and includes auxiliary attributes such as traveled distance, inferred activity at each location (e.g., home, work/study, frequent visit), and demographic indicators. These include age (categorized into 0–24, 25–44, 45–64, 65+, or not disaggregated), gender (male, female), and income level (\(<\euro 10{,}000\), \(\euro 10{,}000\)--\(15{,}000\), and \(>\euro 15{,}000\)). 

Only trips longer than 500 meters are included, and all movements refer to Spanish residents. The data are particularly valuable for characterizing commuting behavior, inter-regional connectivity, and socio-demographic patterns of mobility.

To retrieve OD data, the user can rely on the get\_od\_data() method. An example is shown in Listing 3.

\begin{lstlisting}[language=Python, caption=Example of a code to download and save a parquet file containing the mobility data between March 20 and March 24.]
from pySpainMobility import Mobility

mobility_data = Mobility(version=2, zones='municipalities', start_date='2022-03-20', end_date='2022-03-24', output_directory='./data/')

mobility_data.get_od_data(keep_activity=True)

\end{lstlisting}
The resulting file is saved as a compressed Parquet dataset, enabling scalable analysis with Python data tools.

\subsection{Number of Trips per Person}
In addition to origin–destination data, the Mobility module allows the extraction of records quantifying the number of trips made by individuals residing in a specific zone. Each entry includes the overnight stay zone, the number of individuals belonging to each demographic group, and the count of trips taken (grouped as 0, 1, 2, or more than 2 trips per day).

This dataset supports behavioral mobility profiling and provides insights into activity intensity across populations and regions. It is particularly suited for temporal comparisons, peak detection, and studies on the relationship between mobility and socioeconomic conditions.

The method get\_number\_of\_trips\_data() can be used as shown in Listing 4.

\begin{lstlisting}[language=Python, caption=Example of a code to download and save a Parquet file containing the mobility data between March 20 and March 24.]
from pySpainMobility import Mobility

mobility_data = Mobility(version=2, zones='municipalities', start_date='2022-03-20', end_date='2022-03-24', output_directory='./data/')

mobility_data.get_number_of_trips_data()
\end{lstlisting}
The output is similarly saved in a Parquet file and aligned with the other mobility indicators.

\subsection{Overnight Stays}
The third available dataset focuses on overnight stays and captures, for each date, the location where individuals spend the night, based on their area of residence. Each record identifies the residence zone, the overnight location, and the number of individuals involved. The data reflect temporary relocation behaviors, such as tourism flows, weekend displacements, or prolonged visits.

This dataset provides a complementary angle to the trip-based indicators. Rather than tracking movement events, it captures the spatial distribution of presence during the nighttime, making it ideal for estimating short-term population concentrations, especially during holidays or seasonal periods.

To download the overnight stays dataset, the user can rely on the following command.

\begin{lstlisting}[language=Python, caption=Example of a code to download and save a Parquet file containing the overnight stays at the municipality level between March 20 and March 24.]
from pySpainMobility import Mobility

mobility_data = Mobility(version=2, zones='municipalities', start_date='2022-03-20', end_date='2022-03-24', output_directory='./data/')

mobility_data.get_overnight_stays_data()

\end{lstlisting}
As with other methods, the output is saved to the specified directory and formatted for immediate use in data analysis environments.

\section{Use Case: Income-stratified Mobility Networks and Seasonal Inequality}
\label{sec:use_case}

Urban and regional inequalities are often studied through the spatial distribution of residence, income, employment, or access to services \cite{moro2021mobility, xu2025using, bruno2026dimensions, saif2019public, chetty2017fading}. Mobility adds a relational dimension to this problem. Places are not only unequal because of their internal socioeconomic composition, but also because they are connected to different populations through flows \cite{eagle2010network, gundougdu2019bridging, arnaiz2025structural}. From a network perspective, inequality can therefore be expressed in the structure of mobility itself \cite{iyer2024mobility, xu2025using, brazil2022environmental, nilforoshan2023human, chang2021mobility}.

Despite the relevance of this perspective, empirical analyses of income-stratified mobility networks remain difficult to reproduce at national scale. They require access to large origin-destination matrices, consistent spatial units, demographic attributes, and a processing pipeline capable of transforming raw daily files into comparable network layers. This is precisely the type of workflow that \texttt{pySpainMobility} is designed to support. By standardizing access to the Spanish open mobility data, the package makes it possible to construct income-specific mobility networks with a transparent and replicable procedure, reducing the technical overhead that often separates open data availability from actual scientific use.

The Spanish mobility data provide a particularly useful setting because origin-destination flows are disaggregated by income category. We exploit this feature to represent the national mobility system as a set of income-specific layers defined over the same geography. In this multilayer representation, each income group generates a directed weighted mobility network. Comparing these layers allows us to ask whether income groups participate in a common spatial system or whether they move through distinct destination geographies.

We use this setting to study income-stratified inter-province mobility across four representative weeks in 2023. Seasonality is not treated as a standalone descriptive object, but as a temporal perturbation of the mobility system. If income-based differences were only a matter of volume, seasonal changes would mainly rescale the layers without altering their relative structure. Conversely, persistent differences in concentration, destination ranking, and destination composition would indicate that income groups occupy distinct positions in the mobility network.

The use case is organized around three linked analytical questions. First, we ask whether income groups differ in the concentration of their inter-province destinations. Second, we examine whether income-specific layers share the same destination hierarchy or instead reveal a separation between low-, middle-, and high-income mobility geographies. Third, we measure whether destination provinces receive income compositions that deviate from the national seasonal baseline, indicating destination-level income selectivity. Together, these analyses show how \texttt{pySpainMobility} can be used not only to retrieve official mobility data, but also to operationalize inequality as a network property: a difference in connectivity, destination structure, and social composition of flows.

\subsection{Network construction}
\label{subsec:usecase_network}

We construct the use case from district-level origin-destination flows for four representative weeks in 2023: 16--22 January, 10--16 April, 17--23 July, and 9--15 October. These weeks are used as seasonal snapshots of winter, spring, summer, and autumn, respectively. For each week, \texttt{pySpainMobility} is used to retrieve the corresponding daily mobility files, harmonize the spatial identifiers, and store the processed data in a format suitable for graph construction.

For each season, we define a directed weighted mobility network \(G_t=(V,E_t)\), where \(V\) is the set of districts and an edge \((i,j)\in E_t\) represents trips from origin district \(i\) to destination district \(j\) during season \(t\). Edge weights correspond to the total number of trips observed during the representative week. Self-loops are excluded, so the analysis focuses on mobility between distinct districts. Indeed, the main analysis focuses on inter-province mobility, defined administratively as flows whose origin and destination districts belong to different provinces.

Income groups are represented as layers of the same mobility system. We construct one aggregate layer and three income-specific layers, corresponding to the income categories available in the source data: low income (\(<10{,}000\) euros), middle income (\(10{,}000\)--\(15{,}000\) euros), and high income (\(>15{,}000\) euros). Each income layer is therefore a directed weighted graph defined over the same spatial system, but with edge weights restricted to trips generated by the corresponding income group.

Although the graphs are constructed at the district level, the main empirical results are reported at province level. We aggregate district-to-district flows into province-to-province flows by summing all edge weights whose origin districts belong to the same origin province and whose destination districts belong to the same destination province. This aggregation improves interpretability for national-scale comparison while preserving the income-layer structure of the mobility network. The resulting province-level network is used to analyse inbound destination shares, income-specific destination hierarchies, and destination-level income selectivity.

The full processing pipeline is retained as part of the reproducible workflow. Supplementary tables report the number of nodes, active edges, and total edge weights by season and income layer, together with additional processing checks (Appendix~\ref{app:usecase_processing}). 

\subsection{Measuring Income-stratified Mobility Inequality}
\label{subsec:usecase_measures}

We quantify income-stratified mobility inequality through three complementary network measures. Each measure captures a different aspect of how income layers are organized in the inter-province mobility network: the concentration of destinations, the similarity between income-specific destination hierarchies, and the income selectivity of destination provinces.

First, we measure the concentration of destinations within each income layer. Let \(w_{jgt}^{in}\) denote the total inbound inter-province flow received by destination province \(j\) from income group \(g\) in season \(t\). The destination share of province \(j\) for income group \(g\) is

\[
s_{jgt} = \frac{w_{jgt}^{in}}{\sum_{j} w_{jgt}^{in}}.
\]

We then compute concentration ratios over the ranked distribution of destination shares \cite{hirschman1980national}. For a given income group and season, the concentration ratio \(CR_k(g,t)\) is defined as

\[
CR_k(g,t) = \sum_{j \in \mathcal{T}_k(g,t)} s_{jgt},
\]

where \(\mathcal{T}_k(g,t)\) is the set of the \(k\) destination provinces with the largest inbound shares for income group \(g\) in season \(t\). High values of \(CR_k\) indicate that an income layer concentrates a large fraction of its inter-province mobility in a small number of destinations.

Second, we compare the destination hierarchies of different income layers. For each season, provinces are ranked according to their inbound share within each income group. We then compute Spearman rank correlations \cite{spearman1987proof} between pairs of income layers. If two income groups have a high correlation, they prioritize a similar set of destination provinces. Low correlations indicate that the groups are embedded in different destination hierarchies, even though they move through the same national spatial system.

Third, we measure the spatial reach of each income layer, following the
notion of a characteristic travel distance in human mobility~\cite{gonzalez2008understanding}. For every inter-province origin-destination pair with available geographic matches, we compute the distance between the representative points of the origin and destination districts. Let \(d_{ij}\) be the distance between origin district \(i\) and destination district \(j\), and let \(w_{ijgt}\) be the corresponding number of trips for income group \(g\) in season \(t\). The weighted mean distance of income layer \(g\) in season \(t\) is

\[
\bar{d}_{gt} = \frac{\sum_{(i,j)} w_{ijgt} d_{ij}}{\sum_{(i,j)} w_{ijgt}},
\]

where the sum is taken over inter-province flows with valid geographic coordinates. This measure captures the average distance traveled by flows, weighting each origin-destination pair by its observed trip volume.

Finally, we introduce a destination income selectivity measure. This measure asks whether a province receives an income composition of inbound travelers that differs from the national seasonal composition of inter-province mobility. For each destination province \(j\), we define the income composition of inbound flows as

\[
p_{jgt} = \frac{w_{jgt}^{in}}{\sum_g w_{jgt}^{in}},
\]

and the national seasonal income composition as

\[
p_{gt}^{nat} = \frac{\sum_j w_{jgt}^{in}}{\sum_j \sum_g w_{jgt}^{in}}.
\]

We compare the destination-specific distribution \(p_{jt}\) with the national baseline \(p_t^{nat}\) using Jensen--Shannon divergence~\cite{lin1991divergence}:

\[
JSD_{jt} = JSD(p_{jt}, p_t^{nat}).
\]

A value close to zero indicates that the destination receives an income mix similar to the national seasonal baseline. Higher values indicate stronger destination income selectivity, meaning that the province attracts an inbound mobility composition disproportionately associated with one or more income groups. To characterize the direction of selectivity, we also identify the income group most overrepresented relative to the national baseline.

Finally, we compute a normalized income-mixing entropy~\cite{shannon1948mathematical,theil1972statistical,reardon2002measures} for each destination,

\[
H_{jt} = -\frac{1}{\log 3}\sum_g p_{jgt}\log p_{jgt},
\]

where \(H_{jt}=1\) corresponds to a perfectly balanced distribution across the three income groups, while lower values indicate stronger dominance by one group. In the main text, entropy is used as a complementary measure to the Jensen--Shannon index. Full tables for all concentration ratios, destination selectivity values, and robustness checks are reported in Appendix~\ref{app:concentration_tables} and Appendix~\ref{app:selectivity_tables}.

\subsection{Results}
\label{subsec:usecase_results}

The income-stratified networks reveal that seasonal mobility inequality is not only a matter of different mobility volumes. Income groups differ in how concentrated their destinations are, in how similarly they rank destination provinces, and in the income composition of the provinces they reach. We present these results in three steps, moving from the structure of income-specific layers to the social selectivity of destinations.

\subsubsection{Unequal Destination Concentration and Hierarchy}
\label{subsubsec:destination_concentration_hierarchy}

We first examine whether income groups are connected to the same destination geography. For each season and income layer, we rank destination provinces by inbound inter-province flow. We then summarize this distribution in two ways. First, \(CR_{10}\) measures the share of each income group's mobility captured by its ten largest destination provinces. Higher values indicate a more concentrated mobility layer. Second, Spearman rank correlations compare the destination rankings of different income layers. High correlations indicate that two income groups assign similar relative importance to destination provinces, while low correlations indicate distinct destination hierarchies.

Table~\ref{tab:income_concentration_similarity} shows a clear income gradient in destination concentration. The low-income layer is the most concentrated across its ten largest destinations in every season: its ten largest destination provinces capture between \(73.6\%\) and \(78.3\%\) of inter-province inbound mobility. Middle-income mobility is substantially less concentrated, while high-income mobility occupies an intermediate position in most seasons. This means that low-income inter-province mobility is organized around a narrower set of destinations, rather than being simply a lower-volume version of the other layers.

\begin{table}[ht]
\centering
\caption{
Income-layer concentration and destination-hierarchy similarity in the inter-province mobility network. 
\(CR_{10}\) reports the percentage of each income group's inbound inter-province mobility 
captured by its ten largest destination provinces. Spearman correlations compare 
destination-share rankings between income layers. Bold values identify the most concentrated 
income layer and the strongest hierarchy alignment in each season.
}
\label{tab:income_concentration_similarity}
\renewcommand{\arraystretch}{1.16}
\begin{tabular}{lccc|ccc}
\toprule
\textbf{Season} & \multicolumn{3}{c}{\textbf{\(CR_{10}\) by income layer (\%)}} & \multicolumn{3}{c}{\textbf{Destination-rank similarity}} \\
\cmidrule(lr){2-4}
\cmidrule(lr){5-7}
& \textbf{Low} & \textbf{Middle} & \textbf{High} & \textbf{Low--middle} & \textbf{Low--high} & \textbf{Middle--high} \\
\midrule
Winter & \textbf{78.3} & 53.4 & 63.3 & 0.535 & 0.151 & \textbf{0.787} \\
Spring & \textbf{73.6} & 54.4 & 69.4 & 0.404 & 0.179 & \textbf{0.843} \\
Summer & \textbf{75.5} & 49.6 & 57.0 & 0.566 & 0.228 & \textbf{0.847} \\
Autumn & \textbf{75.6} & 49.3 & 57.8 & 0.548 & 0.202 & \textbf{0.840} \\
\bottomrule
\end{tabular}
\end{table}

The rank correlations show that income layers also differ in the destinations they prioritize. The low- and high-income layers are weakly aligned in all seasons, with correlations between \(0.151\) and \(0.228\). Low- and middle-income rankings are more similar, but only moderately so, ranging from \(0.404\) to \(0.566\). By contrast, middle- and high-income hierarchies are strongly aligned, with correlations above \(0.78\) in every season. The destination structure of the low-income layer is therefore distinct from the rest of the mobility system, whereas middle- and high-income mobility follow much more similar spatial hierarchies.

Together, these results indicate that income stratification is embedded in the structure of the inter-province mobility network. Low-income mobility is both more concentrated and organized around a different destination hierarchy, implying a narrower effective geography of inter-province connections. Full concentration ratios, including \(CR_1\), \(CR_3\), \(CR_5\), and \(CR_{10}\), and complete pairwise similarity values are reported in Appendix~\ref{app:concentration_tables}.

\subsubsection{Unequal Spatial Reach}
\label{subsubsec:spatial_reach}

The previous results show that income layers differ in the destinations they prioritize. We next examine whether they also differ in spatial reach. For each income layer and season, we compute the weighted mean distance of inter-province trips. Distances are calculated between representative points of origin and destination districts for origin-destination pairs with available geographic matches, and are weighted by trip volume. The measure therefore captures the average distance traveled by observed flows, rather than the average distance between active district pairs.

Table~\ref{tab:income_distance} reports weighted mean distances across the four seasonal snapshots. The high-income layer has the greatest spatial reach in every season. In winter, high-income inter-province trips have a weighted mean distance of \(154.6\) km, compared with \(113.5\) km for the middle-income layer and \(98.6\) km for the low-income layer. The same ordering holds in spring, summer, and autumn, with high-income flows consistently traveling farther than middle- and low-income flows.

\begin{table}[ht]
\centering
\caption{
Weighted mean distance of inter-province trips by income layer and season. 
Distances are computed between representative points of origin and destination districts 
for origin-destination pairs with available geographic matches. Values are kilometers. 
Bold values indicate the income layer with the greatest spatial reach in each season.
}
\label{tab:income_distance}
\renewcommand{\arraystretch}{1.59}
\begin{tabular}{lccc}
\toprule
\textbf{Season} & \textbf{Low income} & \textbf{Middle income} & \textbf{High income} \\
\midrule
Winter & 98.6 & 113.5 & \textbf{154.6} \\
Spring & 116.9 & 124.5 & \textbf{179.7} \\
Summer & 110.9 & 127.7 & \textbf{166.9} \\
Autumn & 114.9 & 129.0 & \textbf{168.1} \\
\bottomrule
\end{tabular}
\end{table}

This result adds a second dimension to the inequality pattern. Low-income mobility is not only more concentrated across destinations; it also has a shorter spatial reach. Conversely, the high-income layer combines a broader destination geography with longer trip distances. Income stratification therefore affects both where groups travel and how far their inter-province mobility extends. The same ordering is visible in the weighted medians reported in Appendix~\ref{app:distance}, which reduces the risk that the result is driven only by long-distance tails. Full distance summaries, including median and interquartile distances, together with geographic match-rate checks, are reported in Appendix~\ref{app:distance}.

\subsubsection{Destination Income Selectivity}
\label{subsubsec:destination_selectivity}

The previous results characterize inequality from the perspective of income layers: some groups concentrate their mobility in fewer destinations and follow different destination hierarchies with diverse spatial reaches. We now reverse the perspective and examine destinations themselves. For each season, we first compute the national income composition of inter-province mobility, that is, the share of all inter-province trips generated by the low-, middle-, and high-income groups. We then compute the same income composition separately for the inbound flows received by each destination province. Destination income selectivity is measured as the Jensen--Shannon divergence~\cite{lin1991divergence} between these two distributions. A province is therefore more selective when the income mix of its arrivals differs more strongly from the national seasonal baseline.

\begin{figure}[t]
    \centering
    \safeincludegraphics[width=\linewidth]{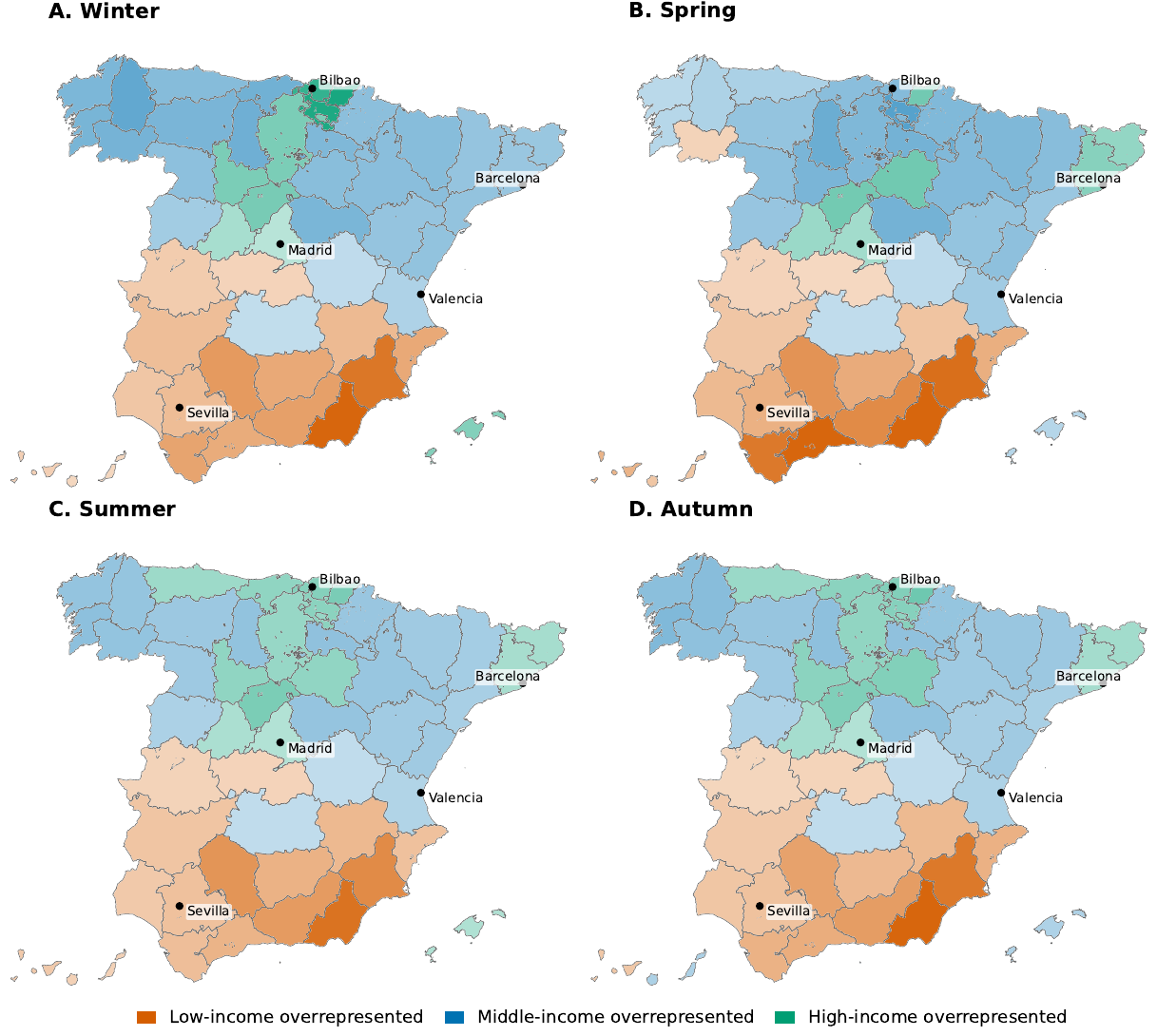}
    \caption{
    Destination income selectivity in Spain's inter-province mobility network.
    Each province is colored according to the income group most overrepresented 
    among inbound inter-province arrivals relative to the national seasonal 
    income composition. Shade intensity is proportional to Jensen--Shannon 
    divergence, with darker shades indicating stronger income selectivity. 
    Black points mark five core cities used as geographic anchors. Island 
    provinces, Ceuta, and Melilla are included in the quantitative analysis but 
    omitted from the map extent for readability.
    }
    \label{fig:income_selectivity_maps}
\end{figure}

Figure~\ref{fig:income_selectivity_maps} maps this measure across seasons. Each province is colored according to the income group most overrepresented among its inbound inter-province arrivals relative to the national seasonal baseline, while shade intensity reflects the magnitude of the Jensen--Shannon divergence. The geography of selectivity is not uniform. Some provinces repeatedly appear as low-income selective destinations, while others attract a disproportionately high-income composition of arrivals. For example, Almeria and Murcia display strong low-income overrepresentation in several seasons, whereas Basque provinces such as Gipuzkoa, Araba/Alava, and Bizkaia stand out as high-income selective in winter.

At the national level, destination income selectivity is stronger in winter and spring than in summer and autumn. The weighted mean Jensen--Shannon divergence is \(0.052\) in winter and \(0.054\) in spring, compared with \(0.035\) in summer and \(0.039\) in autumn. The complementary income-mixing entropy follows the opposite pattern, reaching its highest value in summer. This suggests that the summer expansion of inter-province mobility is associated with a more socially mixed geography of arrivals, whereas winter and spring mobility are more socially selective.

This result adds a destination-level dimension to the multilayer interpretation. Income-stratified mobility inequality is not only visible in the structure of the layers generated by each income group. It is also visible in the social composition of flows received by places. Provinces differ in whether they attract a mobility mix close to the national baseline or a more selective composition dominated by specific income groups. In this sense, destination income selectivity provides a network-based measure of mobility sorting: it identifies where the mobility system connects places to socially distinctive populations.

Full province-level selectivity values, including the most overrepresented income group and income-specific deviations from the national baseline, are reported in Appendix~\ref{app:selectivity_tables}.

\section{Discussion and Conclusion}
\label{subsec:usecase_discussion}

The income-stratified use case shows that \texttt{pySpainMobility} supports more than data retrieval: it enables a complete, reproducible pipeline from official open files to comparable network layers. We discuss the implications along two axes. Section \ref{subsec:usecase_discussion_method} considers the methodological contribution, what the use case demonstrates about the package as an instrument for national-scale, network-based analysis. Section \ref{subsec:usecase_discussion_inequality} turns to the substantive findings on income, mobility, and spatial inequality in Spain.

\subsection{Methodological Contribution}
\label{subsec:usecase_discussion_method}

Beyond its substantive findings, the use case illustrates the methodological value of \texttt{pySpainMobility}. Constructing income-specific mobility layers at national scale requires retrieving large daily origin-destination files across four
seasonal windows, harmonizing district identifiers against the official geometry, filtering by income category, and assembling directed weighted graphs that are comparable across time and groups. Performed manually, this pipeline is error-prone
and difficult to reproduce; the package reduces it to a small, documented set of calls and stores analysis-ready outputs that feed directly into standard network tools.

This reproducibility is what makes the multilayer, multi-season design feasible in the first place. Because access, spatial harmonization, and preprocessing are standardized, the same workflow can be re-run for the other demographic dimensions
available in the data (age and gender), for other spatial aggregations (municipalities or greater urban areas), for other time windows, and for alternative network measures of accessibility, exposure, or segregation. The contribution is therefore not a single analysis but a replicable template: it lowers the technical barrier that usually separates the release of official open mobility data from its substantive scientific use, and it makes national-scale, network-based studies accessible to researchers without a bespoke data-engineering pipeline.

\subsection{Income, Mobility, and Spatial Inequality}
\label{subsec:usecase_discussion_inequality}

The use case shows that income-stratified mobility inequality is not reducible to differences in aggregate trip volume. When income groups are represented as layers of the same inter-province mobility network, inequality becomes visible in the
organization of connections: the low-income layer is more destination-concentrated and shorter-range, low- and high-income layers follow weakly aligned destination hierarchies, and several provinces receive income compositions that deviate from the national seasonal baseline. Income groups thus participate in partially distinct mobility geographies rather than simply moving more or less within a common one.

These findings connect to the literature in the social sciences that treats spatial inequality as a structural property of where groups are located and how they are distributed across space. Research on residential segregation has established both the multiple dimensions along which groups can be separated~\cite{massey1988dimensions, reardon2002measures} and the tight link between rising income inequality and the residential sorting of households by income~\cite{reardon2011income}. Our results echo this literature from a relational angle: the pronounced concentration and shorter reach of the low-income layer,
together with the broader and longer-range geography of high-income mobility, are the flow-based counterpart of the residential ``segregation of affluence''~\cite{reardon2011income}, and our divergence-based selectivity measure adapts the same information-theoretic logic used to quantify residential segregation~\cite{ reardon2002measures}.

At the same time, our analysis contributes to a rapidly growing body of work arguing that segregation is not only a residential condition but also an \emph{experienced} and \emph{relational} one, produced through everyday movement~\cite{cagney2020urban, liao2025socio}. Studies using large-scale mobility data have shown that daily travel connects---or fails to connect---neighborhoods of different socioeconomic  composition~\cite{wang2018urban, phillips2021social}, that
experienced income segregation is embedded in urban mobility
patterns~\cite{moro2021mobility}, that higher-status individuals typically reach more distant and more diverse destinations~\cite{xu2018human}, and that large, dense
cities can paradoxically sustain higher exposure segregation by offering more differentiated  estinations~\cite{nilforoshan2023human}. Our results extend this
agenda in two respects. First, they move the analysis from the intra-urban, single-city, and predominantly US-based settings that dominate this literature to a national, inter-province scale built entirely on official open data. Second, they
show that the weak alignment between low- and high-income destination hierarchies persists across all four seasons: although summer broadens the social mix of arrivals
(higher income-mixing entropy and lower selectivity), it does not dissolve the structural separation between income layers. Seasonality modulates the intensity of mobility sorting without eliminating it.

Limitations should be noted. The income categories are those provided in the official files and are therefore coarse, allowing broad stratification but not a detailed analysis of the income distribution. The analysis uses four representative
weeks rather than a continuous time series, which is appropriate for demonstrating the library and identifying clear patterns but leaves the temporal robustness of the
results for future work. Finally, the measures describe observed
flows, not the mechanisms that generate them: differences between income layers may reflect constraints, preferences, employment structures, family networks, or tourism, which cannot be disentangled from the mobility matrices alone. These
limitations also mark natural extensions---finer socioeconomic strata, longer time series, intra-urban aggregations, and the integration of mobility layers with residential segregation measures---for which the standardized \texttt{pySpainMobility}
workflow provides a direct foundation. Together, these results demonstrate how official open mobility data can be operationalized as a multilayer network for the national-scale study of spatial inequality, reinforcing the broader contribution of \texttt{pySpainMobility}: lowering the technical barrier to mobility analysis while enabling substantive research on networks and forms of inequality.

\section*{List of Abbreviations}

\begin{tabular}{@{}ll@{}}
    \textbf{CDR} & Call Detail Record \\
    \textbf{CR}  & Concentration Ratio (CR$_1$, CR$_3$, CR$_5$, CR$_{10}$) \\
    \textbf{JSD} & Jensen--Shannon Divergence \\
    \textbf{OD}  & Origin--Destination \\
    \textbf{SDK} & Software Development Kit \\
    \textbf{XDR} & eXtended Detail Record \\
\end{tabular}

\section*{Declarations}

\subsection*{Ethics approval and consent to participate}
Not applicable.

\subsection*{Consent for publication}
All authors have read and approved the final manuscript and consent to its publication.
\subsection*{Availability of data and material}
The dataset analysed in this work is publicly available via \url{https://github.com/pySpainMobility/pySpainMobility}.

\subsection*{Competing interests}
The authors declare that they have no competing interests.

\subsection*{Funding}
B.L. has been supported by PRECRISIS (PRotECting public spaces thRough Integrated Smarter Innovative Security), funded by the European Union Internal Security Fund (ISFP-2022-TFI-AG-PROTECT-02-101100539). B.L. and M.L. have been supported by the PNRR ICSC National Research Centre for High Performance Computing, Big Data and Quantum Computing (CN00000013), under the NRRP MUR program funded by the NextGenerationEU. B.L. also acknowledges the support of the PNRR project FAIR - Future AI Research (PE00000013), under the NRRP MUR program funded by the NextGenerationEU. B.L. and M.L. also acknowledge the support of the European Union’s Horizon Europe research and innovation programme under grant agreement No. 101120237 (ELIAS).

\subsection*{Authors' contributions}
C.B. designed and developed the package and performed the experiments. M.L. designed and developed the package and coordinated the writing of the paper. C.B., B.L., T.G.M.R., and M.L. contributed to the interpretation of the results and to writing the manuscript. All authors read and approved the final manuscript.

\subsection*{Acknowledgements}
Not applicable.

\bibliography{sn-bibliography}

@article{luca2023crime,
  title={Crime, inequality and public health: A survey of emerging trends in urban data science},
  author={Luca, Massimiliano and Campedelli, Gian Maria and Centellegher, Simone and Tizzoni, Michele and Lepri, Bruno},
  journal={Frontiers in big Data},
  volume={6},
  pages={1124526},
  year={2023},
  publisher={Frontiers Media SA}
}

@inproceedings{arnaiz2025structural,
  title={Structural Group Unfairness: Measurement and Mitigation by Means of the Effective Resistance},
  author={Arnaiz-Rodriguez, Adrian and Rex, Georgina Curto and Oliver, Nuria},
  booktitle={Proceedings of the International AAAI Conference on Web and Social Media},
  volume={19},
  pages={83--106},
  year={2025}
}

@article{nilforoshan2023human,
  title={Human mobility networks reveal increased segregation in large cities},
  author={Nilforoshan, Hamed and Looi, Wenli and Pierson, Emma and Villanueva, Blanca and Fishman, Nic and Chen, Yiling and Sholar, John and Redbird, Beth and Grusky, David and Leskovec, Jure},
  journal={Nature},
  volume={624},
  number={7992},
  pages={586--592},
  year={2023},
  publisher={Nature Publishing Group UK London}
}

@article{brazil2022environmental,
  title={Environmental inequality in the neighborhood networks of urban mobility in US cities},
  author={Brazil, Noli},
  journal={Proceedings of the National Academy of Sciences},
  volume={119},
  number={17},
  pages={e2117776119},
  year={2022},
  publisher={National Academy of Sciences}
}

@article{gundougdu2019bridging,
  title={The bridging and bonding structures of place-centric networks: Evidence from a developing country},
  author={G{\"u}ndo{\u{g}}du, Didem and Panzarasa, Pietro and Oliver, Nuria and Lepri, Bruno},
  journal={PloS one},
  volume={14},
  number={9},
  pages={e0221148},
  year={2019},
  publisher={Public Library of Science San Francisco, CA USA}
}

@article{eagle2010network,
  title={Network diversity and economic development},
  author={Eagle, Nathan and Macy, Michael and Claxton, Rob},
  journal={Science},
  volume={328},
  number={5981},
  pages={1029--1031},
  year={2010},
  publisher={American Association for the Advancement of Science}
}

@article{chetty2017fading,
  title={The fading American dream: Trends in absolute income mobility since 1940},
  author={Chetty, Raj and Grusky, David and Hell, Maximilian and Hendren, Nathaniel and Manduca, Robert and Narang, Jimmy},
  journal={Science},
  volume={356},
  number={6336},
  pages={398--406},
  year={2017},
  publisher={American Association for the Advancement of Science}
}

@article{saif2019public,
  title={Public transport accessibility: A literature review},
  author={Saif, Muhammad Atiullah and Zefreh, Mohammad Maghrour and Torok, Adam},
  journal={Periodica Polytechnica Transportation Engineering},
  volume={47},
  number={1},
  pages={36--43},
  year={2019}
}

@article{bruno2026dimensions,
  title={The dimensions of accessibility: proximity, opportunities, values},
  author={Bruno, Matteo and Campanelli, Bruno and M. Melo, Hygor P and Rossi Mori, Lavinia and Loreto, Vittorio},
  journal={EPJ Data Science},
  volume={15},
  number={1},
  pages={22},
  year={2026},
  publisher={Springer}
}

@article{kotov2026spanishoddata,
  title={spanishoddata: A package for accessing and working with Spanish Open Mobility Big Data},
  author={Kotov, Egor and Vidal-Tortosa, Eugeni and Cant{\'u}-Ros, Oliva G and Burrieza-Gal{\'a}n, Javier and Herranz, Ricardo and Gull{\'o}n Mu{\~n}oz-Repiso, Tania and Lovelace, Robin},
  journal={Environment and Planning B: Urban Analytics and City Science},
  pages={23998083251415040},
  year={2026},
  publisher={SAGE Publications Sage UK: London, England}
}

@article{xu2025using,
  title={Using human mobility data to quantify experienced urban inequalities},
  author={Xu, Fengli and Wang, Qi and Moro, Esteban and Chen, Lin and Salazar Miranda, Arianna and Gonz{\'a}lez, Marta C and Tizzoni, Michele and Song, Chaoming and Ratti, Carlo and Bettencourt, Luis and others},
  journal={Nature human behaviour},
  volume={9},
  number={4},
  pages={654--664},
  year={2025},
  publisher={Nature Publishing Group UK London}
}

@article{bohorquez2009,
  title={Common ecology quantifies human insurgency},
  author={Bohorquez, J. and Gourley, S. and Dixon, A. and Spagat, M. and Johnson, N.},
  journal={Nature},
  volume={462},
  pages={911--914},
  year={2009}
}

@article{blondel2015survey,
  title={A survey of results on mobile phone datasets analysis},
  author={Blondel, Vincent D and Decuyper, Adeline and Krings, Gautier},
  journal={EPJ data science},
  volume={4},
  number={1},
  pages={10},
  year={2015},
  publisher={Springer}
}

@article{aleta2022quantifying,
  title={Quantifying the importance and location of SARS-CoV-2 transmission events in large metropolitan areas},
  author={Aleta, Alberto and Mart{\'\i}n-Corral, David and Bakker, Michiel A and Pastore y Piontti, Ana and Ajelli, Marco and Litvinova, Maria and Chinazzi, Matteo and Dean, Natalie E and Halloran, M Elizabeth and Longini Jr, Ira M and others},
  journal={Proceedings of the National Academy of Sciences},
  volume={119},
  number={26},
  pages={e2112182119},
  year={2022},
  publisher={National Acad Sciences}
}

@article{chang2021mobility,
  title={Mobility network models of COVID-19 explain inequities and inform reopening},
  author={Chang, Serina and Pierson, Emma and Koh, Pang Wei and Gerardin, Jaline and Redbird, Beth and Grusky, David and Leskovec, Jure},
  journal={Nature},
  volume={589},
  number={7840},
  pages={82--87},
  year={2021},
  publisher={Nature Publishing Group}
}

@article{singh2015money,
  title={Money walks: implicit mobility behavior and financial well-being},
  author={Singh, Vivek Kumar and Bozkaya, Burcin and Pentland, Alex},
  journal={PloS one},
  volume={10},
  number={8},
  pages={e0136628},
  year={2015},
  publisher={Public Library of Science San Francisco, CA USA}
}

@article{tatem2017worldpop,
  title={WorldPop, open data for spatial demography},
  author={Tatem, Andrew J},
  journal={Scientific data},
  volume={4},
  number={1},
  pages={1--4},
  year={2017},
  publisher={Nature Publishing Group}
}

@article{wiedermann2022evidence,
  title={Evidence for positive long-and short-term effects of vaccinations against COVID-19 in wearable sensor metrics--Insights from the German Corona Data Donation Project},
  author={Wiedermann, Marc and Rose, Annika H and Maier, Benjamin F and Kolb, Jakob J and Hinrichs, David and Brockmann, Dirk},
  journal={arXiv preprint arXiv:2204.02846},
  year={2022}
}

@article{beneduce2025urban,
  title={Urban Safety Perception Through the Lens of Large Multimodal Models: A Persona-based Approach},
  author={Beneduce, Ciro and Lepri, Bruno and Luca, Massimiliano},
  journal={arXiv preprint arXiv:2503.00610},
  year={2025}
}

@article{moreira2013predicting,
  title={Predicting taxi--passenger demand using streaming data},
  author={Moreira-Matias, Luis and Gama, Joao and Ferreira, Michel and Mendes-Moreira, Joao and Damas, Luis},
  journal={IEEE Transactions on Intelligent Transportation Systems},
  volume={14},
  number={3},
  pages={1393--1402},
  year={2013},
  publisher={IEEE}
}

@article{pappalardo2021evaluation,
  title={Evaluation of home detection algorithms on mobile phone data using individual-level ground truth},
  author={Pappalardo, Luca and Ferres, Leo and Sacasa, Manuel and Cattuto, Ciro and Bravo, Loreto},
  journal={EPJ data science},
  volume={10},
  number={1},
  pages={29},
  year={2021},
  publisher={Springer Berlin Heidelberg}
}

@article{llorente2015social,
  title={Social media fingerprints of unemployment},
  author={Llorente, Alejandro and Garcia-Herranz, Manuel and Cebrian, Manuel and Moro, Esteban},
  journal={PloS one},
  volume={10},
  number={5},
  pages={e0128692},
  year={2015},
  publisher={Public Library of Science San Francisco, CA USA}
}

@article{luca2022modeling,
  title={Modeling international mobility using roaming cell phone traces during COVID-19 pandemic},
  author={Luca, Massimiliano and Lepri, Bruno and Frias-Martinez, Enrique and Lutu, Andra},
  journal={EPJ Data Science},
  volume={11},
  number={1},
  pages={22},
  year={2022},
  publisher={Springer Berlin Heidelberg}
}

@article{chen2019complete,
	
	Author = {Chen, Guangshuo and Viana, Aline Carneiro and Fiore, Marco and Sarraute, Carlos},
	
	Journal = {EPJ Data Science},
	Number = {1},
	Pages = {30},
	Title = {Complete trajectory reconstruction from sparse mobile phone data},

	Volume = {8},
	Year = {2019},
	}

@article{pappalardo2015returners,
  title={Returners and explorers dichotomy in human mobility},
  author={Pappalardo, Luca and Simini, Filippo and Rinzivillo, Salvatore and Pedreschi, Dino and Giannotti, Fosca and Barab{\'a}si, Albert-L{\'a}szl{\'o}},
  journal={Nature communications},
  volume={6},
  number={1},
  pages={1--8},
  year={2015},
  publisher={Nature Publishing Group}
}

@article{bagrow2011,
  title={Collective response of human populations to large-scale emergencies},
  author={Bagrow, J. P. and Wang, D. and Barabasi, A.-L.},
  journal={PLos One},
  volume={6},
  number={3},
  pages={e17680},
  year={2011}
}

@techreport{MITMA2023,
  title        = {Informe metodol\'ogico del Estudio de Movilidad con Big Data},
  author       = {{Ministerio de Transportes y Movilidad Sostenible}},
  year         = {2023},
  institution  = {Ministerio de Transportes y Movilidad Sostenible},
  address      = {Madrid, Espa\~na},
  url          = {https://www.transportes.gob.es/recursos_mfom/paginabasica/recursos/a3_informe_metodologico_estudio_movilidad_mitms_v8.pdf},
  note         = {Versi\'on 8},
  type         = {Informe t\'ecnico}
}

@article{gullon2022hermes,
  author    = {Gull\'on-Mu\~noz-Repiso, T. and Rodr\'iguez Cano, C.I.},
  title     = {HERMES. Nodo de acceso al Sistema Nacional de Transporte},
  journal   = {Revista del Ministerio de Fomento},
  year      = {2022},
  number    = {722},
  pages     = {64--749},
  issn      = {1577-4589}
}

@article{poncedeleon2022covidflowmaps,
  author    = {Ponce de Le\'on, M. and Gull\'on-Mu\~noz-Repiso, T. and Rodr\'iguez Cano, C.I.},
  title     = {COVID-19 Flow-Maps: an open geographic information system on COVID-19 and human mobility for Spain},
  journal   = {Scientific Data},
  year      = {2022},
  number    = {310},
  note      = {Published 2021}, 
}

@article{gullon2023movilidad,
  author    = {Gull\'on-Mu\~noz-Repiso, T. and Picornell, M.},
  title     = {Movilidad m\'as precisa con tecnolog\'ia bigdata e inteligencia artificial},
  journal   = {Revista MITMA},
  year      = {2023},
  number    = {736},
  pages     = {20--29},
  issn      = {2792-4564},
  month     = {May}
}

@article{conesa2024mixture,
  author    = {Conesa, D. and L\'opez de Rioja, V. and Gull\'on, T. and Tauste Campo, A. and Prats, C. and Alvarez-Lacalle, E. and Echebarria, B.},
  title     = {A mixture of mobility and meteorological data provides a high correlation with COVID-19 growth in an infection-naive population: a study for Spanish provinces},
  journal   = {Frontiers in Public Health},
  year      = {2024},
  volume    = {12},
  pages     = {1288531},
  doi       = {10.3389/fpubh.2024.1288531},
  pmid      = {38528860},
  pmcid     = {PMC10962055},
  month     = {Mar}
}

@article{yabe2024enhancing,
  title={Enhancing human mobility research with open and standardized datasets},
  author={Yabe, Takahiro and Luca, Massimiliano and Tsubouchi, Kota and Lepri, Bruno and Gonzalez, Marta C and Moro, Esteban},
  journal={Nature Computational Science},
  volume={4},
  number={7},
  pages={469--472},
  year={2024},
  publisher={Nature Publishing Group US New York}
}

@article{luca2021survey,
  title={A survey on deep learning for human mobility},
  author={Luca, Massimiliano and Barlacchi, Gianni and Lepri, Bruno and Pappalardo, Luca},
  journal={ACM Computing Surveys (CSUR)},
  volume={55},
  number={1},
  pages={1--44},
  year={2021},
  publisher={ACM New York, NY}
}

@article{moro2021mobility,
  title={Mobility patterns are associated with experienced income segregation in large US cities},
  author={Moro, Esteban and Calacci, Dan and Dong, Xiaowen and Pentland, Alex},
  journal={Nature communications},
  volume={12},
  number={1},
  pages={4633},
  year={2021},
  publisher={Nature Publishing Group UK London}
}

@article{batty2013,
  title={Building a science of cities},
  author={Batty, Michael},
  journal={Cities},
  volume={29},
  pages={S9--S16},
  year={2012},
  publisher={Elsevier}
}

@article{mazzoli2019,
  title={Field theory for recurrent mobility},
  author={Mazzoli, M. and Molas, A. and Bassolas, A. and Lenormand, M. and Colet, P. and Ramasco, J. J.},
  journal={Nature Communications},
  volume={10},
  pages={3895},
  year={2019}
}

@article{moreno2025critical,
  title={Critical mobility in policy making for epidemic containment},
  author={Moreno L{\'o}pez, Jes{\'u}s A and Mateo, David and Hernando, Alberto and Meloni, Sandro and Ramasco, Jos{\'e} J},
  journal={Scientific Reports},
  volume={15},
  number={1},
  pages={3055},
  year={2025},
  publisher={Nature Publishing Group UK London}
}

@article{toth2021inequality,
  title={Inequality is rising where social network segregation interacts with urban topology},
  author={T{\'o}th, Gerg{\H{o}} and Wachs, Johannes and Di Clemente, Riccardo and Jakobi, {\'A}kos and S{\'a}gv{\'a}ri, Bence and Kert{\'e}sz, J{\'a}nos and Lengyel, Bal{\'a}zs},
  journal={Nature communications},
  volume={12},
  number={1},
  pages={1143},
  year={2021},
  publisher={Nature Publishing Group UK London}
}

@article{shannon1948mathematical,
  title={A mathematical theory of communication},
  author={Shannon, Claude Elwood},
  journal={The Bell system technical journal},
  volume={27},
  number={3},
  pages={379--423},
  year={1948},
  publisher={Nokia Bell Labs}
}

@article{theil1972statistical,
  title={Statistical decomposition analysis: With applications in the social and administrative sciences},
  author={Theil, Henri},
  journal={(No Title)},
  year={1972}
}

@article{massey1988dimensions,
  title={The dimensions of residential segregation},
  author={Massey, Douglas S and Denton, Nancy A},
  journal={Social forces},
  volume={67},
  number={2},
  pages={281--315},
  year={1988},
  publisher={The University of North Carolina Press}
}

@article{reardon2011income,
  title={Income inequality and income segregation},
  author={Reardon, Sean F and Bischoff, Kendra},
  journal={American journal of sociology},
  volume={116},
  number={4},
  pages={1092--1153},
  year={2011},
  publisher={University of Chicago Press Chicago, IL}
}

@article{wang2018urban,
  title={Urban mobility and neighborhood isolation in America’s 50 largest cities},
  author={Wang, Qi and Phillips, Nolan Edward and Small, Mario L and Sampson, Robert J},
  journal={Proceedings of the National Academy of Sciences},
  volume={115},
  number={30},
  pages={7735--7740},
  year={2018},
  publisher={National Academy of Sciences}
}

@article{phillips2021social,
  title={The social integration of American cities: Network measures of connectedness based on everyday mobility across neighborhoods},
  author={Phillips, Nolan E and Levy, Brian L and Sampson, Robert J and Small, Mario L and Wang, Ryan Q},
  journal={Sociological Methods \& Research},
  volume={50},
  number={3},
  pages={1110--1149},
  year={2021},
  publisher={SAGE Publications Sage CA: Los Angeles, CA}
}

@article{xu2018human,
  title={Human mobility and socioeconomic status: Analysis of Singapore and Boston},
  author={Xu, Yang and Belyi, Alexander and Bojic, Iva and Ratti, Carlo},
  journal={Computers, Environment and Urban Systems},
  volume={72},
  pages={51--67},
  year={2018},
  publisher={Elsevier}
}

@article{liao2025socio,
  title={Socio-spatial segregation and human mobility: A review of empirical evidence},
  author={Liao, Yuan and Gil, Jorge and Yeh, Sonia and Pereira, Rafael HM and Alessandretti, Laura},
  journal={Computers, Environment and Urban Systems},
  volume={117},
  pages={102250},
  year={2025},
  publisher={Elsevier}
}

@article{cagney2020urban,
  title={Urban mobility and activity space},
  author={Cagney, Kathleen A and York Cornwell, Erin and Goldman, Alyssa W and Cai, Liang},
  journal={Annual Review of Sociology},
  volume={46},
  number={1},
  pages={623--648},
  year={2020},
  publisher={Annual Reviews}
}

@article{reardon2002measures,
  title={Measures of multigroup segregation},
  author={Reardon, Sean F and Firebaugh, Glenn},
  journal={Sociological methodology},
  volume={32},
  number={1},
  pages={33--67},
  year={2002},
  publisher={Wiley Online Library}
}

@article{lin1991divergence,
  title={Divergence measures based on the Shannon entropy},
  author={Lin, Jianhua},
  journal={IEEE Transactions on Information theory},
  volume={37},
  number={1},
  pages={145--151},
  year={1991},
  publisher={IEEE}
}

@article{iyer2024mobility,
  title={Mobility and transit segregation in urban spaces},
  author={Iyer, Nandini and Menezes, Ronaldo and Barbosa, Hugo},
  journal={Environment and Planning B: Urban Analytics and City Science},
  volume={51},
  number={7},
  pages={1496--1512},
  year={2024},
  publisher={SAGE Publications Sage UK: London, England}
}

@article{iyer2025understanding,
  title={Understanding Urban-Rural Disparities in Mobility Inefficiency for Colombia, Mexico, and India},
  author={Iyer, Nandini and Luca, Massimiliano and Di Clemente, Riccardo},
  journal={arXiv preprint arXiv:2503.01810},
  year={2025}
}

@article{yabe2020quantifying,
  title={Quantifying the economic impact of disasters on businesses using human mobility data: a Bayesian causal inference approach},
  author={Yabe, Takahiro and Zhang, Yunchang and Ukkusuri, Satish V},
  journal={EPJ Data Science},
  volume={9},
  number={1},
  pages={36},
  year={2020},
  publisher={Springer Berlin Heidelberg}
}

@article{yabe2022mobile,
  title={Mobile phone location data for disasters: A review from natural hazards and epidemics},
  author={Yabe, Takahiro and Jones, Nicholas KW and Rao, P Suresh C and Gonzalez, Marta C and Ukkusuri, Satish V},
  journal={Computers, Environment and Urban Systems},
  volume={94},
  pages={101777},
  year={2022},
  publisher={Elsevier}
}

@book{hirschman1980national,
  title={National power and the structure of foreign trade},
  author={Hirschman, Albert O},
  volume={105},
  year={1980},
  publisher={Univ of California Press}
}

@article{spearman1987proof,
  title={The proof and measurement of association between two things},
  author={Spearman, Charles},
  journal={The American journal of psychology},
  volume={100},
  number={3/4},
  pages={441--471},
  year={1987},
  publisher={JSTOR}
}

@article{gonzalez2008understanding,
  title={Understanding individual human mobility patterns},
  author={Gonzalez, Marta C and Hidalgo, Cesar A and Barabasi, Albert-Laszlo},
  journal={nature},
  volume={453},
  number={7196},
  pages={779--782},
  year={2008},
  publisher={Nature Publishing Group UK London}
}

@article{park2024post,
  title={Post-disaster recovery policy assessment of urban socio-physical systems},
  author={Park, Sangung and Yabe, Takahiro and Ukkusuri, Satish V},
  journal={Computers, Environment and Urban Systems},
  volume={114},
  pages={102184},
  year={2024},
  publisher={Elsevier}
}

@article{perkins2014theory,
  title={Theory and data for simulating fine-scale human movement in an urban environment},
  author={Perkins, T Alex and Garcia, Andres J and Paz-Sold{\'a}n, Valerie A and Stoddard, Steven T and Reiner Jr, Robert C and Vazquez-Prokopec, Gonzalo and Bisanzio, Donal and Morrison, Amy C and Halsey, Eric S and Kochel, Tadeusz J and others},
  journal={Journal of the Royal Society Interface},
  volume={11},
  number={99},
  pages={20140642},
  year={2014},
  publisher={The Royal Society}
}

@article{colizza2008epidemic,
  title={Epidemic modeling in metapopulation systems with heterogeneous coupling pattern: Theory and simulations},
  author={Colizza, Vittoria and Vespignani, Alessandro},
  journal={Journal of theoretical biology},
  volume={251},
  number={3},
  pages={450--467},
  year={2008},
  publisher={Elsevier}
}

@article{colizza2007modeling,
  title={Modeling the worldwide spread of pandemic influenza: baseline case and containment interventions},
  author={Colizza, Vittoria and Barrat, Alain and Barthelemy, Marc and Valleron, Alain-Jacques and Vespignani, Alessandro},
  journal={PLoS medicine},
  volume={4},
  number={1},
  pages={e13},
  year={2007},
  publisher={Public Library of Science San Francisco, USA}
}

@article{eubank2004modelling,
  title={Modelling disease outbreaks in realistic urban social networks},
  author={Eubank, Stephen and Guclu, Hasan and Anil Kumar, VS and Marathe, Madhav V and Srinivasan, Aravind and Toroczkai, Zoltan and Wang, Nan},
  journal={Nature},
  volume={429},
  number={6988},
  pages={180--184},
  year={2004},
  publisher={Nature Publishing Group}
}
\clearpage
\appendix

\section{Supplementary material}
\label{app:usecase}

This appendix reports supplementary material for the use case presented in 
Section~\ref{sec:use_case}. It documents the graph-construction audit, provides 
full concentration and destination-hierarchy tables, reports supplementary 
destination income-selectivity results, and gives additional statistics for the 
spatial-reach analysis.

\subsection{Processing audit}
\label{app:usecase_processing}

Table~\ref{tab:usecase_processing_audit} summarizes the seasonal graph objects 
created for the use case before the province-level aggregation. The audit 
reports the number of active nodes, active directed edges, and total edge 
weight for each representative week.

\begin{table}[!htbp]
\centering
\caption{
Processing audit for the seasonal mobility graphs. Edge weights correspond to 
the total number of trips observed during each representative week.
}
\label{tab:usecase_processing_audit}
\renewcommand{\arraystretch}{1.15}
\begin{tabular}{lrrr}
\toprule
\textbf{Season} & \textbf{Nodes} & \textbf{Directed edges} & \textbf{Total weight} \\
\midrule
Winter & 3854 & 837839 & 386.0M \\
Spring & 3854 & 1048379 & 390.7M \\
Summer & 3856 & 1123523 & 377.2M \\
Autumn & 3856 & 1107444 & 380.0M \\
\bottomrule
\end{tabular}
\end{table}

A small number of origin-destination identifiers appearing in the official 
mobility files could not be matched to the released district geometry layer. 
Since this inconsistency originates from the official source data, we do not 
alter the origin-destination matrices. We retain the official flows for network 
construction and use geographic attributes only when a match is available.

\FloatBarrier

\subsection{Destination concentration and hierarchy}
\label{app:concentration_tables}

Table~\ref{tab:concentration_ratios_full} extends the concentration results 
reported in Table~\ref{tab:income_concentration_similarity}. It reports 
\(CR_1\), \(CR_3\), \(CR_5\), and \(CR_{10}\), where \(CR_k\) is the percentage 
of each income group's inbound inter-province mobility captured by its top 
\(k\) destination provinces.

\begin{table}[!htbp]
\centering
\caption{
Concentration ratios of inter-province destination shares by income layer and 
season. Values are percentages.
}
\label{tab:concentration_ratios_full}
\renewcommand{\arraystretch}{1.12}
\begin{tabular}{llrrrr}
\toprule
\textbf{Season} & \textbf{Income layer} & \(\mathbf{CR_1}\) & \(\mathbf{CR_3}\) & \(\mathbf{CR_5}\) & \(\mathbf{CR_{10}}\) \\
\midrule
Winter & Low & 15.1 & 35.9 & 51.6 & 78.3 \\
Winter & Middle & 14.2 & 27.5 & 36.5 & 53.4 \\
Winter & High & 19.9 & 35.7 & 47.4 & 63.3 \\
\addlinespace[0.25em]
Spring & Low & 13.4 & 32.5 & 47.6 & 73.6 \\
Spring & Middle & 14.2 & 28.3 & 37.3 & 54.4 \\
Spring & High & 26.9 & 47.1 & 54.6 & 69.4 \\
\addlinespace[0.25em]
Summer & Low & 15.4 & 34.5 & 48.5 & 75.5 \\
Summer & Middle & 12.4 & 24.3 & 33.1 & 49.6 \\
Summer & High & 19.6 & 33.9 & 41.4 & 57.0 \\
\addlinespace[0.25em]
Autumn & Low & 14.8 & 34.7 & 48.5 & 75.6 \\
Autumn & Middle & 13.0 & 25.5 & 33.8 & 49.3 \\
Autumn & High & 20.7 & 35.0 & 43.0 & 57.8 \\
\bottomrule
\end{tabular}
\end{table}

Table~\ref{tab:income_similarity_full} reports the full pairwise Spearman rank 
correlations between income-specific destination hierarchies. These values 
correspond to the similarity block in Table~\ref{tab:income_concentration_similarity}.

\begin{table}[!htbp]
\centering
\caption{
Spearman rank correlations between income-specific destination hierarchies by 
season.
}
\label{tab:income_similarity_full}
\renewcommand{\arraystretch}{1.12}
\begin{tabular}{lrrr}
\toprule
\textbf{Season} & \textbf{Low--middle} & \textbf{Low--high} & \textbf{Middle--high} \\
\midrule
Winter & 0.535 & 0.151 & 0.787 \\
Spring & 0.404 & 0.179 & 0.843 \\
Summer & 0.566 & 0.228 & 0.847 \\
Autumn & 0.548 & 0.202 & 0.840 \\
\bottomrule
\end{tabular}
\end{table}

\FloatBarrier

\subsection{Destination income selectivity}
\label{app:selectivity_tables}

Table~\ref{tab:income_selectivity_summary_app} reports the national seasonal 
summary of destination income selectivity and income mixing. The weighted mean 
Jensen-Shannon divergence summarizes how strongly destination provinces deviate 
from the national seasonal income composition of inter-province mobility. The 
weighted mean entropy summarizes the income mixing of destination provinces.

\begin{table}[!htbp]
\centering
\caption{
National summary of destination income selectivity and income mixing by season.
}
\label{tab:income_selectivity_summary_app}
\renewcommand{\arraystretch}{1.12}
\begin{tabular}{lrrr}
\toprule
\textbf{Season} & \textbf{Weighted mean JSD} & \textbf{Weighted mean entropy} & \textbf{Destinations} \\
\midrule
Winter & 0.052 & 0.703 & 52 \\
Spring & 0.054 & 0.715 & 52 \\
Summer & 0.035 & 0.742 & 52 \\
Autumn & 0.039 & 0.738 & 52 \\
\bottomrule
\end{tabular}
\end{table}

Tables~\ref{tab:top10_low_income_destinations}--\ref{tab:top10_high_income_destinations}
report the destination provinces where each income layer is most 
overrepresented relative to the national seasonal baseline. Provinces are 
ranked by overrepresentation, measured as the difference between the income 
group's share among arrivals to the destination and the same income group's 
national share of inter-province mobility. The JSD column reports the overall 
income selectivity of the destination province.

\FloatBarrier

\begingroup
\footnotesize
\setlength{\tabcolsep}{4pt}
\renewcommand{\arraystretch}{0.96}

\begin{longtable}{@{}lrp{3.7cm}rrrr@{}}
\caption{Top 10 destination provinces by low-income overrepresentation.}
\label{tab:top10_low_income_destinations}\\
\toprule
\textbf{Season} & \textbf{Rank} & \textbf{Province} & \textbf{JSD} & \textbf{Dest.} & \textbf{Nat.} & \textbf{Overrep.} \\
& & & & \textbf{(\%)} & \textbf{(\%)} & \textbf{(pp)} \\
\midrule
\endfirsthead

\caption[]{Top 10 destination provinces by low-income overrepresentation (continued).}\\
\toprule
\textbf{Season} & \textbf{Rank} & \textbf{Province} & \textbf{JSD} & \textbf{Dest.} & \textbf{Nat.} & \textbf{Overrep.} \\
& & & & \textbf{(\%)} & \textbf{(\%)} & \textbf{(pp)} \\
\midrule
\endhead

\midrule
\multicolumn{7}{r}{Continued on next page}\\
\endfoot

\bottomrule
\endlastfoot

Winter & 1 & Almeria & 0.207 & 67.2 & 17.6 & 49.7 \\
 & 2 & Murcia & 0.132 & 55.7 & 17.6 & 38.1 \\
 & 3 & Cordoba & 0.085 & 46.3 & 17.6 & 28.7 \\
 & 4 & Granada & 0.076 & 45.3 & 17.6 & 27.8 \\
 & 5 & Cadiz & 0.071 & 43.2 & 17.6 & 25.6 \\
 & 6 & Jaen & 0.064 & 41.4 & 17.6 & 23.9 \\
 & 7 & Alicante/Alacant & 0.064 & 41.3 & 17.6 & 23.7 \\
 & 8 & Ceuta & 0.063 & 41.3 & 17.6 & 23.7 \\
 & 9 & Malaga & 0.061 & 40.6 & 17.6 & 23.0 \\
 & 10 & Sevilla & 0.038 & 36.8 & 17.6 & 19.3 \\
\addlinespace[0.35em]

Spring & 1 & Ceuta & 0.194 & 69.3 & 19.6 & 49.7 \\
 & 2 & Malaga & 0.181 & 67.9 & 19.6 & 48.2 \\
 & 3 & Almeria & 0.188 & 66.6 & 19.6 & 46.9 \\
 & 4 & Cadiz & 0.129 & 59.9 & 19.6 & 40.2 \\
 & 5 & Murcia & 0.143 & 58.4 & 19.6 & 38.8 \\
 & 6 & Granada & 0.105 & 53.5 & 19.6 & 33.9 \\
 & 7 & Cordoba & 0.093 & 52.5 & 19.6 & 32.9 \\
 & 8 & Sevilla & 0.054 & 44.9 & 19.6 & 25.3 \\
 & 9 & Jaen & 0.061 & 44.1 & 19.6 & 24.4 \\
 & 10 & Santa Cruz de Tenerife & 0.049 & 42.1 & 19.6 & 22.5 \\
\addlinespace[0.35em]

Summer & 1 & Almeria & 0.138 & 53.3 & 14.0 & 39.3 \\
 & 2 & Murcia & 0.105 & 46.9 & 14.0 & 32.9 \\
 & 3 & Granada & 0.083 & 41.5 & 14.0 & 27.5 \\
 & 4 & Cordoba & 0.092 & 41.0 & 14.0 & 27.0 \\
 & 5 & Malaga & 0.050 & 34.1 & 14.0 & 20.1 \\
 & 6 & Ceuta & 0.053 & 33.3 & 14.0 & 19.3 \\
 & 7 & Jaen & 0.052 & 33.2 & 14.0 & 19.2 \\
 & 8 & Cadiz & 0.035 & 30.7 & 14.0 & 16.7 \\
 & 9 & Alicante/Alacant & 0.031 & 30.2 & 14.0 & 16.2 \\
 & 10 & Albacete & 0.043 & 29.3 & 14.0 & 15.3 \\
\addlinespace[0.35em]

Autumn & 1 & Almeria & 0.157 & 56.1 & 14.4 & 41.7 \\
 & 2 & Murcia & 0.130 & 51.8 & 14.4 & 37.4 \\
 & 3 & Ceuta & 0.083 & 42.2 & 14.4 & 27.7 \\
 & 4 & Granada & 0.081 & 41.5 & 14.4 & 27.1 \\
 & 5 & Cordoba & 0.073 & 39.4 & 14.4 & 25.0 \\
 & 6 & Malaga & 0.059 & 36.6 & 14.4 & 22.2 \\
 & 7 & Alicante/Alacant & 0.053 & 36.3 & 14.4 & 21.9 \\
 & 8 & Cadiz & 0.052 & 35.1 & 14.4 & 20.7 \\
 & 9 & Jaen & 0.042 & 32.6 & 14.4 & 18.2 \\
 & 10 & Sevilla & 0.031 & 29.7 & 14.4 & 15.3 \\

\end{longtable}
\endgroup

\begingroup
\footnotesize
\setlength{\tabcolsep}{4pt}
\renewcommand{\arraystretch}{0.96}

\begin{longtable}{@{}lrp{3.7cm}rrrr@{}}
\caption{Top 10 destination provinces by middle-income overrepresentation.}
\label{tab:top10_middle_income_destinations}\\
\toprule
\textbf{Season} & \textbf{Rank} & \textbf{Province} & \textbf{JSD} & \textbf{Dest.} & \textbf{Nat.} & \textbf{Overrep.} \\
& & & & \textbf{(\%)} & \textbf{(\%)} & \textbf{(pp)} \\
\midrule
\endfirsthead

\caption[]{Top 10 destination provinces by middle-income overrepresentation (continued).}\\
\toprule
\textbf{Season} & \textbf{Rank} & \textbf{Province} & \textbf{JSD} & \textbf{Dest.} & \textbf{Nat.} & \textbf{Overrep.} \\
& & & & \textbf{(\%)} & \textbf{(\%)} & \textbf{(pp)} \\
\midrule
\endhead

\midrule
\multicolumn{7}{r}{Continued on next page}\\
\endfoot

\bottomrule
\endlastfoot

Winter & 1 & Lugo & 0.082 & 87.3 & 63.4 & 23.9 \\
 & 2 & Pontevedra & 0.063 & 86.1 & 63.4 & 22.8 \\
 & 3 & Ourense & 0.065 & 85.6 & 63.4 & 22.2 \\
 & 4 & A Coruna & 0.060 & 83.5 & 63.4 & 20.1 \\
 & 5 & Castellon/Castello & 0.041 & 83.5 & 63.4 & 20.1 \\
 & 6 & Tarragona & 0.040 & 81.2 & 63.4 & 17.8 \\
 & 7 & La Rioja & 0.057 & 80.7 & 63.4 & 17.3 \\
 & 8 & Zamora & 0.045 & 80.1 & 63.4 & 16.7 \\
 & 9 & Guadalajara & 0.061 & 80.0 & 63.4 & 16.6 \\
 & 10 & Leon & 0.061 & 79.9 & 63.4 & 16.5 \\
\addlinespace[0.35em]

Spring & 1 & Araba/Alava & 0.092 & 89.4 & 61.7 & 27.7 \\
 & 2 & Palencia & 0.074 & 86.6 & 61.7 & 24.9 \\
 & 3 & Valladolid & 0.062 & 84.4 & 61.7 & 22.7 \\
 & 4 & La Rioja & 0.066 & 83.4 & 61.7 & 21.7 \\
 & 5 & Zamora & 0.045 & 82.5 & 61.7 & 20.8 \\
 & 6 & Castellon/Castello & 0.044 & 82.4 & 61.7 & 20.7 \\
 & 7 & Leon & 0.039 & 80.2 & 61.7 & 18.5 \\
 & 8 & Cantabria & 0.048 & 79.8 & 61.7 & 18.1 \\
 & 9 & Burgos & 0.057 & 79.1 & 61.7 & 17.4 \\
 & 10 & Salamanca & 0.037 & 78.8 & 61.7 & 17.1 \\
\addlinespace[0.35em]

Summer & 1 & Pontevedra & 0.045 & 79.0 & 62.4 & 16.7 \\
 & 2 & Castellon/Castello & 0.029 & 78.5 & 62.4 & 16.1 \\
 & 3 & Ourense & 0.040 & 78.5 & 62.4 & 16.1 \\
 & 4 & La Rioja & 0.043 & 78.1 & 62.4 & 15.7 \\
 & 5 & A Coruna & 0.039 & 76.7 & 62.4 & 14.4 \\
 & 6 & Lugo & 0.038 & 76.2 & 62.4 & 13.9 \\
 & 7 & Zaragoza & 0.036 & 73.4 & 62.4 & 11.0 \\
 & 8 & Huesca & 0.029 & 73.4 & 62.4 & 11.0 \\
 & 9 & Valencia/Valencia & 0.014 & 73.4 & 62.4 & 11.0 \\
 & 10 & Guadalajara & 0.039 & 73.3 & 62.4 & 10.9 \\
\addlinespace[0.35em]

Autumn & 1 & Pontevedra & 0.056 & 83.4 & 62.0 & 21.4 \\
 & 2 & A Coruna & 0.047 & 79.5 & 62.0 & 17.5 \\
 & 3 & Castellon/Castello & 0.029 & 79.3 & 62.0 & 17.3 \\
 & 4 & Ourense & 0.044 & 79.0 & 62.0 & 17.0 \\
 & 5 & Lugo & 0.049 & 78.9 & 62.0 & 16.9 \\
 & 6 & La Rioja & 0.043 & 77.0 & 62.0 & 15.0 \\
 & 7 & Guadalajara & 0.042 & 75.3 & 62.0 & 13.3 \\
 & 8 & Valencia/Valencia & 0.018 & 75.1 & 62.0 & 13.1 \\
 & 9 & Salamanca & 0.019 & 71.8 & 62.0 & 9.8 \\
 & 10 & Huesca & 0.035 & 71.3 & 62.0 & 9.3 \\

\end{longtable}
\endgroup

\begingroup
\footnotesize
\setlength{\tabcolsep}{4pt}
\renewcommand{\arraystretch}{0.96}

\begin{longtable}{@{}lrp{3.7cm}rrrr@{}}
\caption{Top 10 destination provinces by high-income overrepresentation.}
\label{tab:top10_high_income_destinations}\\
\toprule
\textbf{Season} & \textbf{Rank} & \textbf{Province} & \textbf{JSD} & \textbf{Dest.} & \textbf{Nat.} & \textbf{Overrep.} \\
& & & & \textbf{(\%)} & \textbf{(\%)} & \textbf{(pp)} \\
\midrule
\endfirsthead

\caption[]{Top 10 destination provinces by high-income overrepresentation (continued).}\\
\toprule
\textbf{Season} & \textbf{Rank} & \textbf{Province} & \textbf{JSD} & \textbf{Dest.} & \textbf{Nat.} & \textbf{Overrep.} \\
& & & & \textbf{(\%)} & \textbf{(\%)} & \textbf{(pp)} \\
\midrule
\endhead

\midrule
\multicolumn{7}{r}{Continued on next page}\\
\endfoot

\bottomrule
\endlastfoot

Winter & 1 & Gipuzkoa & 0.145 & 55.0 & 19.1 & 35.9 \\
 & 2 & Araba/Alava & 0.140 & 54.6 & 19.1 & 35.5 \\
 & 3 & Bizkaia & 0.121 & 48.8 & 19.1 & 29.8 \\
 & 4 & Segovia & 0.064 & 39.4 & 19.1 & 20.3 \\
 & 5 & Valladolid & 0.060 & 33.0 & 19.1 & 13.9 \\
 & 6 & Illes Balears & 0.054 & 31.8 & 19.1 & 12.7 \\
 & 7 & Burgos & 0.061 & 31.7 & 19.1 & 12.7 \\
 & 8 & Avila & 0.023 & 27.7 & 19.1 & 8.7 \\
 & 9 & Madrid & 0.008 & 26.2 & 19.1 & 7.1 \\
 & 10 & Palencia & 0.071 & 25.4 & 19.1 & 6.3 \\
\addlinespace[0.35em]

Spring & 1 & Segovia & 0.067 & 38.9 & 18.7 & 20.3 \\
 & 2 & Soria & 0.070 & 38.5 & 18.7 & 19.8 \\
 & 3 & Gipuzkoa & 0.068 & 36.5 & 18.7 & 17.9 \\
 & 4 & Barcelona & 0.050 & 33.2 & 18.7 & 14.5 \\
 & 5 & Madrid & 0.027 & 33.0 & 18.7 & 14.3 \\
 & 6 & Avila & 0.034 & 31.1 & 18.7 & 12.4 \\
 & 7 & Girona & 0.040 & 28.4 & 18.7 & 9.7 \\
 & 8 & Zaragoza & 0.054 & 24.2 & 18.7 & 5.5 \\
 & 9 & Teruel & 0.039 & 23.4 & 18.7 & 4.7 \\
 & 10 & Navarra & 0.056 & 23.2 & 18.7 & 4.5 \\
\addlinespace[0.35em]

Summer & 1 & Segovia & 0.061 & 46.2 & 23.6 & 22.5 \\
 & 2 & Gipuzkoa & 0.062 & 43.5 & 23.6 & 19.9 \\
 & 3 & Soria & 0.041 & 39.1 & 23.6 & 15.4 \\
 & 4 & Bizkaia & 0.052 & 36.7 & 23.6 & 13.1 \\
 & 5 & Valladolid & 0.043 & 35.8 & 23.6 & 12.2 \\
 & 6 & Cantabria & 0.042 & 35.1 & 23.6 & 11.4 \\
 & 7 & Madrid & 0.014 & 34.4 & 23.6 & 10.8 \\
 & 8 & Girona & 0.024 & 34.1 & 23.6 & 10.4 \\
 & 9 & Avila & 0.020 & 33.5 & 23.6 & 9.8 \\
 & 10 & Illes Balears & 0.016 & 33.1 & 23.6 & 9.4 \\
\addlinespace[0.35em]

Autumn & 1 & Segovia & 0.056 & 45.0 & 23.6 & 21.4 \\
 & 2 & Gipuzkoa & 0.071 & 44.9 & 23.6 & 21.3 \\
 & 3 & Soria & 0.054 & 43.0 & 23.6 & 19.4 \\
 & 4 & Bizkaia & 0.055 & 36.2 & 23.6 & 12.6 \\
 & 5 & Valladolid & 0.047 & 35.9 & 23.6 & 12.3 \\
 & 6 & Avila & 0.024 & 35.4 & 23.6 & 11.8 \\
 & 7 & Madrid & 0.013 & 34.4 & 23.6 & 10.8 \\
 & 8 & Girona & 0.027 & 34.2 & 23.6 & 10.7 \\
 & 9 & Cantabria & 0.047 & 33.5 & 23.6 & 9.9 \\
 & 10 & Burgos & 0.042 & 32.4 & 23.6 & 8.8 \\

\end{longtable}
\endgroup

\FloatBarrier

\subsection{Spatial reach by income layer}
\label{app:distance}

Table~\ref{tab:distance_full_app} reports supplementary distance statistics 
for the inter-province flows analysed in Section~\ref{subsubsec:spatial_reach}. 
Distances are computed between representative points of origin and destination 
districts for origin-destination pairs with available geographic matches. All 
statistics are weighted by trip volume, so they describe the distance 
distribution of observed mobility flows rather than the unweighted distribution 
of active origin-destination pairs. The table uses only the inter-province 
subset of flows, defined as trips whose origin and destination districts belong 
to different provinces.

The supplementary statistics confirm the pattern reported in the main text. In 
every season, the high-income layer has the largest weighted mean and median 
distance, while the low-income layer has the shortest spatial reach. The 
interquartile range also shows that high-income mobility has a longer upper 
tail, especially in spring, summer, and autumn.

\begin{table}[!htbp]
\centering
\caption{
Weighted distance statistics for inter-province trips by income layer and 
season. Values are kilometers. Mean, median, and interquartile distances are 
weighted by trip volume.
}
\label{tab:distance_full_app}
\renewcommand{\arraystretch}{1.12}
\begin{tabular}{llrrrr}
\toprule
\textbf{Season} & \textbf{Income layer} & \textbf{Mean} & \textbf{Median} & \textbf{P25} & \textbf{P75} \\
\midrule
Winter & Low income & 98.6 & 44.1 & 20.9 & 98.4 \\
Winter & Middle income & 113.5 & 61.1 & 25.9 & 116.7 \\
Winter & High income & 154.6 & 79.4 & 44.2 & 170.3 \\
Spring & Low income & 116.9 & 61.6 & 25.0 & 117.9 \\
Spring & Middle income & 124.5 & 68.1 & 29.9 & 136.1 \\
Spring & High income & 179.7 & 98.5 & 58.3 & 256.5 \\
Summer & Low income & 110.9 & 56.5 & 22.7 & 114.9 \\
Summer & Middle income & 127.7 & 71.3 & 30.2 & 139.1 \\
Summer & High income & 166.9 & 90.9 & 49.5 & 231.6 \\
Autumn & Low income & 114.9 & 58.4 & 23.6 & 117.9 \\
Autumn & Middle income & 129.0 & 73.6 & 32.1 & 143.5 \\
Autumn & High income & 168.1 & 95.6 & 53.9 & 236.8 \\
\bottomrule
\end{tabular}
\end{table}

\FloatBarrier

\end{document}